\documentclass{article}
\usepackage{spconf,amsmath,graphicx}
\usepackage{tabularx}
\usepackage{makecell}
\usepackage[section]{placeins} 
\usepackage{adjustbox}
\usepackage{tablefootnote}
\usepackage{multirow}
\usepackage{siunitx}
\usepackage{arydshln}
\usepackage{mathtools}
\usepackage{fancyhdr}
\usepackage{enumitem}
\setlist{nosep, leftmargin=14pt}

\usepackage{mwe} 

\title{Prediction of recurrence free survival of head and neck cancer using PET/CT radiomics and clinical information}

\name{Mona Furukawa \textsuperscript{1,2}, \qquad Daniel R. McGowan\textsuperscript{3,4}, \qquad Bart\l omiej W. Papie\.z\textsuperscript{1,2}}

\address{\textsuperscript{1}  Big Data Institute, Li Ka Shing Centre for Health Information and Discovery,
\\ University of Oxford, Old Road Campus, Oxford, OX3 7LF, UK \\
    \textsuperscript{2} Nuffield Department of Population Health, 
     \\University of Oxford, Old Road Campus, Oxford, OX3 7LF, UK\\
     \textsuperscript{3} Department of Oncology, University of Oxford, Old Road Campus Research Building
\\Roosevelt Drive, Oxford, OX3 7DQ\\
  \textsuperscript{4} Department of Medical Physics and Clinical Engineering, 
  \\Oxford University Hospitals NHS FT, Churchill Hospital, Oxford, OX3 7LE
}

\fancypagestyle{firstpage}{%

\fancyfoot[C]{For the purpose of open access, the author has applied a Creative Commons Attribution (CC BY) license to any Author Accepted Manuscript version arising.}
}

\begin{document}
\thispagestyle{firstpage}

\maketitle
\begin{abstract}
The 5-year survival rate of Head and Neck Cancer (HNC) has not improved over the past decade and one common cause of treatment failure is recurrence. In this paper, we built Cox proportional hazard (CoxPH) models that predict the recurrence free survival (RFS) of oropharyngeal HNC patients. Our models utilise both clinical information and multimodal radiomics features extracted from tumour regions in Computed Tomography (CT) and Positron Emission Tomography (PET). 
Furthermore, we were one of the first studies to explore the impact of segmentation accuracy on the predictive power of the extracted radiomics features, through under- and over-segmentation study. Our models were trained using the HEad and neCK TumOR (HECKTOR) challenge data, and the best performing model achieved a concordance index (C-index) of 0.74 for the model utilising clinical information and multimodal CT and PET radiomics features, which compares favourably with the model that only used clinical information (C-index of 0.67). Our under- and over-segmentation study confirms that segmentation accuracy affects radiomics extraction, however, it affects PET and CT differently.

\end{abstract}
\section{Introduction}
\label{sec:intro}

\subsection{Head and neck cancer}
Head and neck cancer (HNC) is the seventh most prevalent cancer in the world. Approximately 660,000 people are diagnosed with the disease annually \cite{gormley2022reviewing}. Common causes of treatment failure and death in HNC is distance metastases and second primary cancers. Survival and treatment outcome of patients can potentially be improved by identifying prognostic factors that indicate tumour aggressiveness or risk of recurrence during diagnosis. In doing so, patients can be stratified into different risk groups and, this can be taken into account when chemotherapy regimens and radiation doses are being planned for patients \cite{vallieres2017radiomics}.

\subsection{Medical imaging}
Medical imaging such as 2-deoxy-2-[$\prescript{18}{}{\mathbf{F}}$] fluoro-D-glucose (FDG) Positron Emission Tomography (PET) and Computed Tomography (CT) contain large amounts of quantitative mineable information including clinical information, such as prognostic features. \textit{Radiomics} is the process of extracting quantitative information from images, and using them with an aim to aid clinical decision making \cite{vallieres2017radiomics,rizzo2018radiomics,aerts2014decoding}.

\subsection{Radiomics}
Radiomics features are extracted from the segmented region of interest (e.g. tumour) of the image \cite{radiomicsimg}. These features are hypothesised to have prognostic power and, analysis by \cite{aerts2014decoding} found that some radiomics features have associations with gene-expression pattern in tumours. Furthermore, \cite{bogowicz2017comparison} have also observed that CT and PET based radiomics have capabilities to predict the control rate of the local tumour in HNC. These examples demonstrate a wide range of predictive power radiomics hold. There are different types of radiomics features, such as shape, first-order and second-order statistics. Shape features represent information such as the volume and sphericity of the tumour. First-order statistics describe histogram-based properties of the image such as the skewness. Lastly, second-order statistic features provide textural information such as tumour heterogeneity \cite{vallieres2017radiomics,rizzo2018radiomics}.  

\subsection{Aims}
This study aims to create multimodal model that can predict the recurrence free survival (RFS) of patients diagnosed with HNC (located in the oropharynx region). 
Our contributions are as follows. 
First, a novel model was built using multimodal radiomics extracted from FDG-PET/CT and non-imaging data (clinical information) \cite{HECKTOR2022}. 
Second, this study also explored the effects of segmentation accuracy on the quality of the radiomics features extracted for RFS prediction to understand the importance of segmentation accuracy in the multimodal RFS workflow \cite{rizzo2018radiomics,HECKTOR2022}. 
We were one of the first studies to explore the impact of segmentation accuracy on the predictive power of the extracted multimodal radiomics features for RFS prediction.

\section{Data and Methods}
\label{sec:format}
\subsection{Data}
\label{ssec:subhead}
The HEad and neCK TumOR (HECKTOR) challenge \cite{HECKTOR2022} provided FDG-PET and CT images of patients diagnosed with oropharyngeal HNC. Furthermore, for each patient, the PET images were registered with CT images. In addition to image data, clinical information - centre ID, gender, age, weight, tobacco, alcohol consumption, performance status, HPV status, surgery and/or chemotherapy were provided for majority of the patients. The data included in the challenge was built from 7 different centres with 524 subjects. 

Moreover, the training dataset images were provided with the ground truth labels for primary Gross Tumour Volumes (GTVp) and metastatic lymph nodes (GTVn). A subset of the cases ($n=488$) contained RFS information. As this study aims to build RFS prediction models, this subset of dataset was used for the study. This training dataset was further split into training (85 \%), validation (7.5 \%) and testing (7.5 \%) sets. Further information about the challenge dataset can be found in \cite{HECKTOR2022}. 

\subsection{Preprocessing}

\begin{figure}[htb]
  \centering
  \includegraphics[width=7cm]{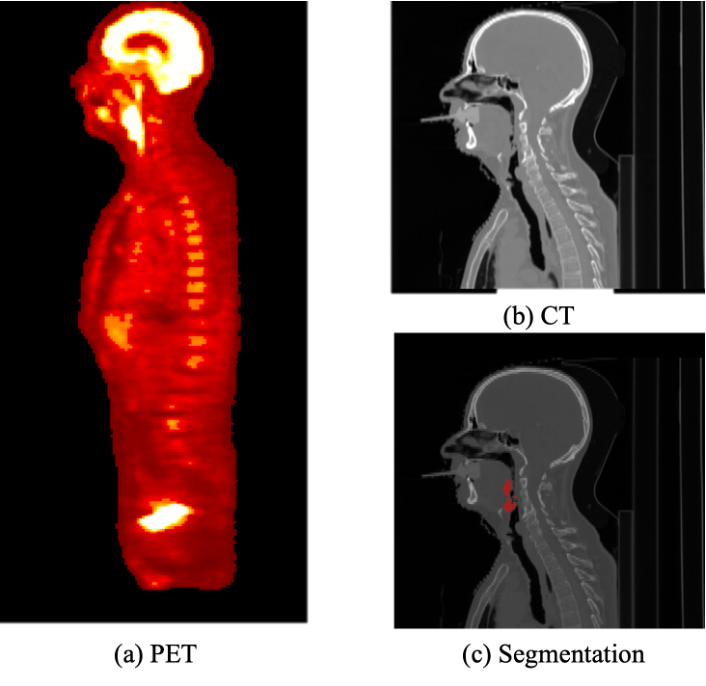}
  \caption{An exemplar (a) PET (b) CT and (c) Segmentation 
  of a patient from HECKTOR challenge \cite{HECKTOR2022}}\medskip
  \label{PET/CT}
\end{figure}

The clinical dataset was pre-processed for survival analysis as some participants had missing values. For binary data, (tobacco, alcohol, HPV status and surgery), missing values were set as zero, negative status, denoted as 0, was replaced with -1 and positive status, denoted as 1, was kept as it is. The missing data for weight was set as 75 kg as instructed by HECKTOR 2022 and performance status and centre ID were dropped from the dataset \cite{HECKTOR2022,wanghecktor2022}. 

PET, CT, and their annotations (segmentation labels) were resampled so that they have the same isotropic voxel spacing of 2 mm $\times$ 2 mm $\times$ 2 mm. During this process, Bspline and nearest neighbour interpolator were used for PET/CT and annotations respectively. Furthermore, the field of view of PET, CT and their annotations were different for some patients (e.g. Fig.~\ref{PET/CT}). To ensure uniformity in size, PET, CT and labels were cropped using bounding boxes. The bounding boxes were determined by finding the field of view that covers both PET and CT for each patient using code from \cite{HECKTOR2022}. Lastly, the annotation of GTVp and GTVn were merged as a single label, so the radiomics features from GTVp and GTVn could be extracted together.

\begin{figure*}
  \centering
  \includegraphics[width=17.8cm]{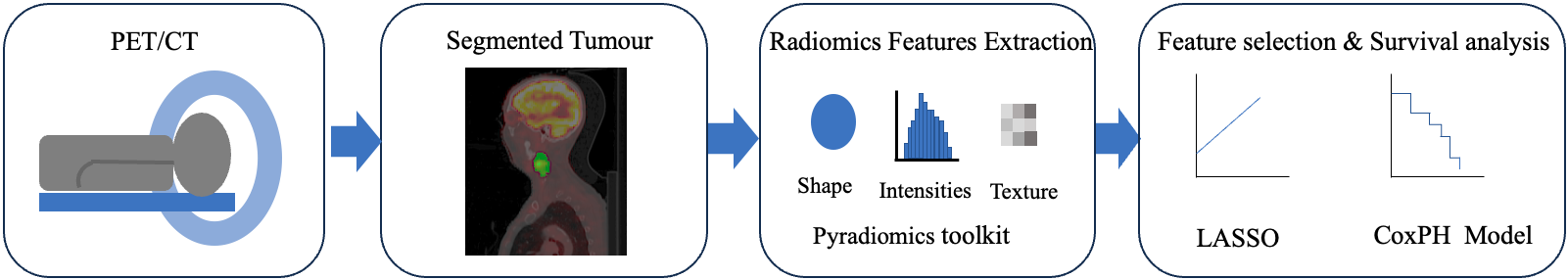}
  \caption{Radiomics workflow (Image adapted from \cite{radiomicsimg} and method adapted from \cite{bourigault2021multimodal})}
  \label{radWorkflow}
\end{figure*}

\subsection{Radiomics workflow}

The radiomics workflow (shown in Fig.~\ref{radWorkflow}) typically involves, image acquisition, image segmentation, extraction of radiomics features, feature selection, and model fitting using the extracted features \cite{radiomicsimg}. The pyradiomics package \cite{pyradiomics} was used to extract features from the tumour regions in PET and CT (107 features each). After feature extraction, the radiomics features of PET and CT were compared, so that features with different values could be relabelled as PET- and CT-specific features. Features that had the same value (e.g. volume) were considered common features. 
In the next step, features were selected using the least absolute shrinkage and selection operator (LASSO)\cite{LASSO} with 8-fold cross-validation. The selected features were then fed into the Cox proportional hazard (CoxPH) model \cite{bourigault2021multimodal,kleinbaum1996survival}.

\subsubsection{Cox proportional hazard model}
Cox proportional hazard (CoxPH) models are used to model the effects of an explanatory variable on survival. The mathematical expression of the CoxPH model is shown in Eq.~\eqref{coxPHmodel}, and it expresses the hazard at time $t$ for an individual that have the set of explanatory variables $\{\mathbf{X}|\mathbf{X}=(X_1,X_2,...X_p)\}$. $h_{0}(t)$ is the baseline hazard function and, the coefficient ($\beta_i$) of the explanatory variables are estimated by maximising the likelihood function \cite{kleinbaum1996survival}. The CoxPH model was used in this study to model the RFS, which is the time without any recurrence measured in days after the end of treatment \cite{HECKTOR2022}. 

Here, the dataset was used to evaluate the predictive power of the model, which was evaluated by calculating the concordance index (C-index). The C-index takes a value between 0 and 1, where 1 indicates a perfect concordance \cite{Davidson-Pilon2019}. Lastly, different combinations of imaging and non-imaging features were fed into the LASSO and CoxPH analysis to find the optimal combinations of features for the prediction model \cite{bourigault2021multimodal}. 

\begin{equation}
h(t,\mathbf{X})=h_{0}(t)\exp{\left(\sum_{i=1}^p{\beta_iX_i}\right)}
\label{coxPHmodel}
\end{equation}

\subsection{Impact of segmentation quality}
As shown in Fig.~\ref{radWorkflow}, segmentation is one of the crucial processes in the radiomics workflow as features will be extracted from the region of interest (e.g. tumour area) \cite{radiomicsimg}. Therefore, to study how the accuracy of segmentation affects the quality of the radiomics features, morphological operations of the ground truth annotations were performed to study over-segmentation (using dilation), and under-segmentation (using erosion).
By changing the extend of dilation/erosion (defined by $r$ as the radius of structuring element in voxel units), the resulting segmentations were used to extract the radiomics features and fed into the LASSO and CoxPH analysis.
In this study, we consider the segmentation accuracy to be represented by the deviation (erosion/dilation) annotations from the ground truth, while the quality of the radiomics features is represented by the C-index value on RFS.

\section{Results}
\label{sec:typestyle}

The CoxPH regression model for predicting RFS was built using different combinations of radiomics features and clinical information and the results are shown in Tab.~\ref{result}. The best performing model achieved a C-index of 0.74, and this was built using both PET and CT features and clinical information. Of all the ground truth models, PET+clinical model had the lowest performance, C-index of 0.65. This was even lower than the clinical only model.

One example of radiomics features extracted and selected for the CoxPH regression model was sphericity, which describes the roundness of the tumour compared to a sphere. Another example was gray level size zone matrix small area emphasis. This can be used to understand how much fine textures there are within the tumour \cite{pyradiomics}.

The result from the under-segmentation and over-segmen\-tation study is shown in Tab.~\ref{result} from the third row onward. As shown, the eroded CT+clinical model had a comparable performance to the the ground truth CT model, which suggests that the CT radiomics features were not affected by under-segmentation. However, CT+clinical model was more sensitive to over-segmentation. On the other hand, PET+clinical model performance was affected by under-segmentation and had the best performance when annotations were over-segmented by $r$=2 (with C-index of 0.68).

Lastly, the volume of some tumours was too small when the ground truth annotations were eroded by $r$=2 or more, so radiomics features (both for PET and CT) could not be extracted, therefore no results are presented.

\begin{table}[htb]
\resizebox{1\columnwidth}{!}{
\begin{tabular}{c|c|c}
\textbf{CoxPH Regression Model} & \textbf{C-index (Train)} & \textbf{C-index (Test)}\\
\hline
Clinical & 0.63 & 0.67\\
\hline
Ground truth (CT $+$ Clinical) & 0.77 & 0.73\\
Ground truth (PET $+$ Clinical) & 0.74 &0.65 \\
Ground truth (CT$+$PET$+$Clinical) & 0.78 & \textbf{0.74} \\
\hline
Eroded r=1 (CT $+$ Clinical)  & 0.76 & 0.74\\
Eroded r=1(PET $+$ Clinical) & 0.73 &0.61\\
Eroded r=1(CT$+$PET$+$Clinical) & 0.77 & 0.69\\
\hline
Dilated r=1 (CT$+$ Clinical)  & 0.75 & 0.68 \\
Dilated r=1 (PET $+$ Clinical)  &0.72 & 0.62 \\
Dilated r=1 (CT$+$PET$+$Clinical) &0.76 &0.64\\
\hline
Dilated r=2 (CT$+$Clinical) &0.75 &0.68\\
Dilated r=2 (PET$+$Clinical)&0.73&0.68\\
Dilated r=2 (CT$+$PET$+$Clinical)&0.77&0.67\\
\hline
\end{tabular}
}
\caption{RFS prediction using different combination of imaging (derived from CT and/or PET) and non-imaging (clinical) features. Ground truth, eroded (under-segmentation) and dilated (over-segmentation) means that the radiomics features were extracted using the ground truth annotations, eroded, and dilated annotations, respectively.} 
\label{result}
\end{table}

\section{Discussion}
\label{sec:majhead}
CoxPH model built using PET/CT and clinical features had the best performance (C-index of 0.74). This performance was better than the clinical only model (C-index of 0.67), which shows that additional information from radiomics features improved the prediction accuracy of RFS. However, it is important to note that PET+clinical model performed the worst out of all the ground truth models. This suggests that PET radiomics features may not be contributing substantially to the survival predictions. This can potentially be explained by a couple of reasons. 

The first potential reason is that the bin width used to discretise the images to extract radiomics. Image discretisation is done before feature extraction and the default bin width equal to 25 provided by pyradiomics \cite{pyradiomics} was used in this study. However, it is important to remember that PET and CT have a different range of measurement, and PET has a smaller range than CT, which means that bin width could have been too large for PET. If this was the case, we may have lost some information when extracting PET radiomics features \cite{radiomicsimg}. This will be further explored and fine-tuned for future extraction process. 

Another potential reason is that many of the ground truth annotations were initially drawn on the CT of the PET/CT scan. This means that the ground truth annotations may not accurately cover the tumours in PET \cite{HECKTOR2022}. Interestingly, this may be true considering how PET radiomics model performed better when annotations were dilated by r=2 (C-index of 0.68). This result may suggest that the area around the tumour boundary may contain useful information for survival prediction. This can be further supported by the fact that out of all the PET models, the eroded model had the lowest performance (C-index of 0.61). Normally, the area around the tumour boundary is quite blurry in PET due to its low resolution and many existing segmentation methods often ignore the blurred edges \cite{li2019variational}. This result suggests that the information from PET may contain valuable information and should be considered when developing a segmentation algorithm for radiomics analysis.  

Furthermore, it is interesting to see that unlike PET features, CT features were not affected by under-segmentation. However, their performance dropped for over-segmentation. These preliminary findings suggest that in order to build a successful multimodal prediction model using radiomics features, it may be better to have a separate segmentation for PET and CT images \cite{steyaert2023multimodal}. Further investigation should be done to understand why over- and under- segmentation affects CT and PET features differently. One potential way to explore this, is to find selected features that change in value due to under- and over- segmentation. This can also give us a better insight into understanding which features are more sensitive to segmentation accuracy \cite{radiomicsimg}. Overall, the over- and under-segmentation study showed that quality of the radiomics features are affected by the segmentation accuracy. 


\section{Conclusion}
\label{sec:print}

This study highlights two promising key results. Firstly, it has shown that the inclusion of multimodal radiomics features improves the efficacy of the CoxPH model in predicting RFS. Secondly, it has shown that segmentation accuracy does affect the quality of the radiomics features extracted. For the first result, it will be interesting to see if the addition of radiomics features - handcrafted (e.g. number of tumour masses) \cite{rebaud2022simplicity} and deep learning features \cite{bourigault2021multimodal} will help improve the model further. Moreover, it will be intriguing to build a model using only PET or CT radiomics, so that it can be used understand how well image data is able to predict RFS. For the second result, the study has shown that the impact of over- and under-segmentation differs depending on whether the features are extracted from PET or CT. Further study should be done to understand if having a separate segmentation algorithm for PET and CT is better than having one segmentation for both. This has important consequences for multi-modal imaging, as it highlights the importance of tailoring segmentation algorithms depending on the type of medical images. Lastly, further study needs to be done to improve our understanding on how exactly segmentation accuracy affects the radiomics features. Introducing errors that reflect the inter- and intra-observer segmentation variability seen in real settings will help us give a better insight into the implications of segmentation accuracy in radiomics studies \cite{radiomicsimg}. This can be implemented by introducing wide range of segmentation errors, such as inclusion of reader's bias and omission errors \cite{vuadineanu2022analysis}.


\section{Compliance with ethical standards}
\label{sec:ethics}

This research study was conducted retrospectively using
human subject data made available in open access by (HECKTO\-R2022) \cite{HECKTOR2022}. Ethical approval was not required as confirmed by the license attached with the open access data.

\section{Acknowledgments}
\label{sec:acknowledgments}
This work was supported by EPSRC grant number EP/S0240-93/1, GE HealthCare and Centre for Doctoral Training in Sustainable Approaches to Biomedical Science: Responsible and Reproducible Research (SABS : $R^3$) Doctoral Training Centre, University of Oxford.

\bibliographystyle{Mona_ISBI_latex}
\bibliography{Mona_ISBI_latex}

\end{document}